\documentclass[pra,twocolumn,showpacs,superscriptaddress,notitlepage,nofootinbib,amssymb,preprintnumbers]{revtex4-1}

\usepackage{amsmath,amssymb,amsthm}
\usepackage{braket}
\usepackage{empheq}

\usepackage[pdfstartview=FitH]{hyperref}
\hypersetup{
    colorlinks=true,       
    linkcolor=cyan,          
    citecolor=magenta,        
    filecolor=magenta,      
    urlcolor=cyan,           
    runcolor=cyan
}

\newcommand{\ignore}[1]{}

\newcommand{\bes} {\begin{subequations}}
\newcommand{\ees} {\end{subequations}}
\newcommand{\beq}{\begin{equation}}
\newcommand{\eeq}{\end{equation}}
\newcommand{\ba}{\begin{eqnarray}}
\newcommand{\ea}{\end{eqnarray}}

\newcommand{\mrp}{\mathrm{p}}

\newcommand\Tr{\mathrm{Tr}}
\newcommand{\ketbra}[1]{|{#1}\rangle\langle#1|}

\newtheorem*{theorem*}{Theorem}

\newcommand{\mc}{\mathcal}

\def\a{\alpha}
\def\b{\beta}
\def\g{\gamma}

\def\e{\epsilon}

\def\r{\rho}

\def\o{\omega}

\begin{document}
\title{Arbitrary-time error suppression for Markovian adiabatic quantum computing using stabilizer subspace codes}
	
\author{Daniel A. Lidar}
\affiliation{Departments of Electrical \& Computer Engineering, Chemistry, Physics \& Astronomy\\
Center for Quantum Information Science \& Technology\\
University of Southern California, Los Angeles, California 90089, USA}
	
\begin{abstract}
Adiabatic quantum computing (AQC) can be protected against thermal excitations via an encoding into error detecting codes, supplemented with an energy penalty formed from a sum of commuting Hamiltonian terms. 
Earlier work showed that it is possible to suppress the \emph{initial} thermally induced excitation out of the encoded ground state, in the case of local Markovian environments, by using an energy penalty strength that grows only logarithmically in the system size, at a fixed temperature. The question of whether this result applies beyond the initial time was left open.
Here we answer this in the affirmative. We show that thermal excitations out of the encoded ground state can be suppressed at \emph{arbitrary} times under the additional assumption that the total evolution time is polynomial in the system size. Thus, computational problems that can be solved efficiently using AQC in a closed system setting, can still be solved efficiently subject to coupling to a thermal environment. Our construction uses stabilizer subspace codes, which require at least $4$-local interactions to achieve this result. 

\end{abstract}
\maketitle
	
\section{Introduction}	

In adiabatic quantum computing (AQC), computations are performed using a time-dependent Hamiltonian that evolves smoothly from an initial Hamiltonian with a known and easily preparable ground state, to a final Hamiltonian whose ground state is unknown and encodes the desired result~\cite{farhi_quantum_2000} (for a review see Ref.~\cite{Albash-Lidar:RMP}). This model appears promising for near-future large scale realization, especially in terms of (non-universal) quantum annealing devices, which already feature a few thousand qubits~\cite{Harris:2018aa,King:2018aa}. 

Despite enjoying a certain degree of inherent robustness to errors, AQC has four main and well documented failure modes~\cite{childs_robustness_2001,PhysRevLett.95.250503,Aberg:2005rt,ashhab:052330,PhysRevA.75.062313,TAQC,amin_decoherence_2009,PhysRevA.80.022303,Qiang:13,Sarovar:2013kx}: (i) diabatic transitions out of the ground state arising from an evolution on a timescale that is faster than that set by the inverse gap, (ii) control errors resulting in the implementation of the wrong final Hamiltonian, (iii) decoherence of the ground state, and (iv) thermal excitations out of the ground state. 
The first of these is a purely unitary error mode which arises even in the absence of coupling to the environment. It is mitigated by slowing the evolution down in accordance with the adiabatic theorem~\cite{Kato:50}, i.e., in order to remain in the ground state throughout, the total evolution time is required to be large relative to the timescale set by (a small power of) the inverse of the smallest energy gap from the ground state encountered along the evolution~\cite{Jansen:07}. 
The second can be viewed as arising from technical imperfections or from the environment; either way it can be mitigated to some extent by imposing smooth boundary conditions on the interpolation between the initial and final Hamiltonians~\cite{lidar:102106,Wiebe:12,Ge:2015wo,Venuti:2018aa} or by encoding the final Hamiltonian~\cite{Young:2013fk}, but the absence of a complete theory of fault tolerance in AQC (despite impressive attempts~\cite{Mizel:2014sp}) means that it is not currently known how to scalably and reliably overcome control errors. 
The third and fourth are entirely environment-induced errors. Decoherence of the ground state (due to decoherence in the computational basis) is a catastrophic failure mode that occurs when the coupling to the environment is too strong for AQC to be meaningfully executed. Quantum error correction methods can be deployed in principle, but at present they are impractical in that they require the use of many-body interactions that scale with the problem size~\cite{Young:13}. To avoid decoherence of the ground state, AQC should be performed in systems obeying the weak coupling limit to the environment, where decoherence occurs in the instantaneous energy eigenbasis~\cite{Albash:2015nx}. 

In this work we address thermal excitations. This failure mode can be suppressed using a scheme first proposed by Jordan, Farhi, and Shor (JFS)~\cite{jordan2006error}. In the JFS scheme, an error detecting stabilizer subspace code is chosen, and the system Hamiltonian is encoded using the logical operators of the same code. A penalty Hamiltonian $H_{\mrp}$ proportional to the sum of the stabilizer generators of the code is added, which suppresses excitations out of the code subspace. This is useful since without encoding thermal excitations are suppressed only by the gap of the system Hamiltonian, but with encoding, thermal excitations are suppressed by the gap of the penalty Hamiltonian (which is a constant for stabilizer codes) times the magnitude $\eta_{\mrp}$ of the energy penalty. 

In their analysis, JFS assumed a particular system of spins weakly coupled to a photon bath and a pure initial state. They then identified the lowest-weight possible subspace stabilizer codes for detecting $1$-local and $2$-local noise compatible with the suppression of thermal excitation errors. Ref.~\cite{Marvian:2017aa} generalized the JFS suppression result to arbitrary Markovian master equations and arbitrary subspace (as opposed to subsystem) error detection codes, while allowing for mixed initial states.  However, both Refs.~\cite{jordan2006error,Marvian:2017aa} only considered the ultra-short-time performance of this error suppression scheme for Markovian environments. More precisely, they established conditions for the success of the scheme only in terms of the \emph{initial} thermal excitation rate out of the code subspace. 

Here we complete the analysis initiated in Ref.~\cite{Marvian:2017aa} and consider the performance of subspace-based error suppression schemes for \emph{arbitrary} times $t$. We prove that thermal excitation errors can be suppressed for all physically reasonable, local Markovian environments by increasing $\eta_{\mrp}$ only logarithmically in the number of qubits $n$ at constant bath temperature, provided the total evolution time scales at most polynomially in $n$. Our main technical result is formulated in terms of an upper bound on the excited state population at arbitrary $t$, assuming that the system is initialized in the ground subspace. We show that, provided the conditions mentioned above hold, this bound can be made arbitrarily small by increasing $\eta_{\mrp}$ in proportion to $\log(n)$. Since we require that the total evolution time $t_f\sim\mathrm{poly}(n)$, our result does not guarantee protection against thermal excitation errors for problems with exponentially (or superpolynomially) small gaps, for which, by the adiabatic theorem, we expect $t_f$ to have to scale faster than $\mathrm{poly}(n)$. 

The structure of this paper is as follows.
In Sec.~\ref{sec:2} we provide a general bound on the excitation rate out of the ground subspace at arbitrary time. We observe that the bound involves an off-diagonal component (coherence between the ground and excited subspaces) that did not appear in the earlier initial-time treatment of Refs.~\cite{jordan2006error,Marvian:2017aa}. 
In  Sec.~\ref{sec:3} we derive an upper bound on the excited state population after encoding using a subspace-based error detection code and adding an energy penalty term, and show that it can be made arbitrarily small provided the penalty strength $\eta_{\mrp}\sim\log(n)$ and total evolution time $t_f\sim\mathrm{poly}(n)$. This is the content of our main result, Eq.~\eqref{eq:pop-perp-final}. Readers interested primarily in the conclusions can skip many details of the derivation and read the paper starting from this point. We provide a summary and discussion in Sec.~\ref{sec:conc}, and provide a few additional technical details in the Appendix.

\section{Bounding the excitation rate out of the ground subspace at arbitrary time}
\label{sec:2}

Consider the spectral decomposition of a time-dependent Hamiltonian $H(t)$:
\beq
\label{eq:H}
H(t) = \sum_{l\geq 0} \e_l(t) \Pi_l(t)\ ,
\eeq
where $\Pi_l(t)$ denotes the projection onto the (possibly degenerate) $H(t)$-eigensubspace with eigenvalue $\epsilon_l(t)$. The eigenvalues are ordered so that $\e_0(t) \leq \e_1(t) \leq \dots$ $\forall t$, and we assume that there are no level crossings. The eigenprojectors are orthogonal: $\Pi_l(t) \Pi_{l'}(t) = \delta_{ll'}\Pi_l(t)$. From now on we usually drop the explicit time-dependence to simplify the notation. But it important to remember that all our quantities are explicitly time-dependent unless explicitly stated otherwise.

\subsection{General expression for the excitation rate out of the ground subspace}

Assume that the system is initially prepared in the (possibly degenerate) ground subspace of $H$, with energy $\e_{0}$, i.e., $\rho(0)=\Pi_0 \rho(0) \Pi_0$.
Using $\rho = (\Pi_0+\Pi_0^\perp)\rho(\Pi_0+\Pi_0^\perp)$, the population in the subspace orthogonal to $\Pi_0$ is
\begin{align}
p_\perp &\equiv \Tr (\Pi_0^\perp\rho\Pi_0^\perp) = 1-\Tr(\Pi_0\rho\Pi_0)
\end{align}
so that 
\beq
\dot{p}_\perp = -\partial_t\Tr(\Pi_0{\rho}) \equiv -R(t)
\eeq
and
\beq
p_\perp(t) = -\int_0^t R(t')dt' + p_\perp(0)\ .
\label{eq:p-perp}
\eeq
Since we assumed that the initial population is fully in $\Pi_0$, i.e., $\Tr[\Pi_0 \rho(0) \Pi_0]=1$, it follows that $p_\perp(0)=0$.

In contrast to Ref.~\cite{Marvian:2017aa}, which focused on the initial excitation rate out of the ground subspace, $R(0) = \partial_t\Tr(\Pi_0{\rho}) |_{t \simeq 0 }$, here we are interested in the excitation rate for arbitrary $t$ 
\beq
R(t)=\partial_t\Tr(\Pi_0{\rho}) = \Tr(\dot{\Pi}_0{\rho}) + \Tr({\Pi_0}\dot{\rho})\ . 
\label{eq:R(t)}
\eeq
%
Again using $\rho = (\Pi_0+\Pi_0^\perp)\rho(\Pi_0+\Pi_0^\perp)$, we have
\bes
\begin{align}
 \label{eq:A2a}
 \Tr(\dot\Pi_0 {\rho}) &= \Tr(\Pi_0\dot{\Pi}_0\Pi_0\rho) +\Tr(\Pi_0^\perp\dot{\Pi}_0\Pi_0^\perp\rho) \\
 &+ \Tr(\dot{\Pi}_0\Pi_0^\perp\rho\Pi_0) +\Tr(\dot{\Pi}_0 \Pi_0\rho\Pi_0^\perp) \ .
 \label{eq:A2b}
\end{align}
\ees
The terms in line~\eqref{eq:A2a} vanish: differentiating the identity $\Pi_0^2 = \Pi_0$ yields
\beq
\Pi_0 \dot{\Pi}_0 + \dot\Pi_0 \Pi_0= \dot\Pi_0\ \Longrightarrow \ \dot\Pi_0 \Pi_0 = \Pi_0^\perp \dot\Pi_0\ .
\eeq
Multiplying from the left or from the right gives
\beq
\Pi_0\dot\Pi_0 \Pi_0 = 0 \ , \quad \Pi_0^\perp\dot{\Pi}_0\Pi_0^\perp = 0\ .
\eeq

The two complex conjugate terms in line~\eqref{eq:A2b} are due to coherence between $\Pi_0$ and $\Pi_0^\perp$.  
They vanish if we assume that at arbitrary time $t$ the state is in $\Pi_0\oplus\Pi_0^\perp$:
\beq
R(t)= \Tr({\Pi_0}\dot{\rho})\ \ \ \mathrm{if}\ \ \ \rho= \Pi_0 \rho \Pi_0 +\Pi_0^\perp \rho \Pi_0^\perp\ \forall t .
\label{eq:R}
\eeq
Since the initial state satisfies $\rho(0) = \Pi_0 \rho(0) \Pi_0$, 
Eq.~\eqref{eq:R} holds at $t=0$ without an additional no-coherence assumption. This was the case studied in Ref.~\cite{Marvian:2017aa}. But since here we are interested in arbitrary $t$, we have:
\beq
R(t)= \Tr({\Pi_0}\dot{\rho}) + [\Tr(\dot{\Pi}_0 \Pi_0\rho\Pi_0^\perp) + \textrm{c.c.}]. 
\label{eq:R2}
\eeq

One way for the assumption $\rho= \Pi_0 \rho \Pi_0 +\Pi_0^\perp \rho \Pi_0^\perp\ \forall t$ to hold is if decoherence between $\Pi_0$ and $\Pi_0^\perp$ is fast on the timescale of the evolution, i.e., $T_2 \ll t_f$, where $T_2$ is the timescale over which $\Pi_0 \rho \Pi_0^\perp$ and $\Pi_0^\perp \rho \Pi_0$ decay, and $t_f$ is the final time, i.e., the total evolution time. This is certainly true in the time-independent case (where we expect these coherences to decay at least as fast as $e^{-t/T_2}$), but it does not hold in the general time-dependent case, as we discuss below.

\subsection{Adiabatic master equation in Davies-Lindblad form}
\label{sec:AME}

Let us define our open system model. 
Assuming a total Hamiltonian of the form
\beq
H_{\mathrm{tot}}(t) = H_S(t) + H_B + H_{SB} \ ,
\label{eq:Htot}
\eeq
where $H_S(t)$ is the 
time-dependent system Hamiltonian, $H_B$ is a general bath Hamiltonian, and 
\beq
H_{SB}=\sum_{\alpha}{A_{\alpha} \otimes B_{\alpha}}
\eeq
is a general system-bath interaction Hamiltonian, an adiabatic Markovian master equation in Davies-Lindblad form~\cite{Davies:74,Lindblad:76} can be derived in the weak coupling limit~\cite{ABLZ:12-SI}:
\begin{eqnarray}
	\dot{\rho} = \mathcal{L}(t)[\rho] = -i [H(t)+H_{\mathrm{LS}}(t),\rho]+D(t)[\rho]\ ,
	\label{eq:ME}
\end{eqnarray}
where $H_{\mathrm{LS}}(t)$ is the Lamb shift, which commutes with $H_S(t)$, and $D(t)$ denotes the dissipative (non-unitary) part.  Henceforth we use units such that $\hbar\equiv 1$. We assume that the system operators $A_{\alpha}$ are $\ell$-local, with $\ell$ a constant that is independent of the number of system particles (e.g., qubits) $n$. The interaction Hamiltonian $H_{SB}$ then has a local structure, and can be expressed as a sum over $\binom{n}{\ell}$ terms, which is polynomial in $n$.

We briefly review the structure of $D(t)$ (see, e.g., Ref.~\cite{Albash:2015nx} for more details). 
Let $\rho_B$ denote the initial state of the bath. The bath correlation function is
\beq
\langle \mathcal{B}_{\a\a'}(t)\rangle = \Tr(\rho_B e^{-i H_B t} B_\alpha e^{i H_B t} B_{\a'})\ ,
\eeq
and its Fourier transform is
\beq
\gamma_{\alpha\a'}(\omega)= \int_{-\infty}^{\infty}dt\ e^{i\omega t} \langle \mathcal{B}_{\a\a'}(t)\rangle = \g^*_{\a'\a}(\o)\ .
\eeq 
The matrix $\g = \{\gamma_{\alpha\a'}\}$ is positive semi-definite. Therefore it can be diagonalized by a unitary matrix $u$. Define new system operators 
\beq
F_{\a} = \sum_{\a'} (u^\dag)_{\a\a'} A_{\a'}\ ,
\label{eq:F}
\eeq
and their transforms 
\beq
F_{\alpha}(\omega)=\sum_{\epsilon_{l'}-\epsilon_l=\omega}{\Pi_l F_{\alpha}\Pi_{l'}}\ ,
\eeq
where the sum is over all pairs of eigenvalues $\epsilon_{l'}$ and $\epsilon_l$ whose difference is equal to the given Bohr frequency $\o$.
Then, the dissipator $D(t)$ can be written as:
%
\beq
\label{eq:diss}
	D[\rho]=\sum_{\o\alpha}\gamma_{\alpha}(\omega)[F_{\alpha}(\omega)\rho F_{\alpha}^{\dagger}(\omega) -\frac{1}{2}\{F_{\alpha}^{\dagger}(\omega)F_{\alpha}(\omega),\rho\}] \ ,
\eeq
where $\{\g_{\a}(\o)\}$ are the eigenvalues of $\g(\o)$. 

If the bath is in thermal equilibrium at inverse temperature $\b=1/(k_B T)$, then 
the matrix of decay rates satisfies the  Kubo-Martin-Schwinger (KMS) condition~\cite{KMS}:
\begin{eqnarray}
\g_{\alpha}(-\o) = e^{-\b \o}\g_{\alpha}(\o)\ , \quad \o > 0 \ .
\label{eq:KMS}
\end{eqnarray}
It relates the excitation rate $\g_{\alpha}(-\o)$ to the relaxation rate $\g_{\alpha}(\o)$, and shows that excitation is exponentially suppressed in $\b \o$ relative to relaxation.

\subsection{Upper bound on the off-diagonal term}
\label{sec:offdiag-bound}

To bound the off-diagonal term $\Tr(\dot{\Pi}_0 \Pi_0\rho\Pi_0^\perp)$ in Eq.~\eqref{eq:R2}, we note that 
$\dot{\Pi}_0$ can be replaced by the reduced resolvent $S = \Pi_0^\perp(H-\epsilon_0)^{-1}\Pi_0^\perp$. Namely, it is well known that $\dot{\Pi}_0\Pi_0 = -S\dot{H}\Pi_0$  and $\| S\|\leq 1/\Delta(H)$, the inverse minimum spectral gap of $H(t)$, i.e., the gap between $\Pi_0$ and the first excited state (see, e.g., Appendices B and F of Ref.~\cite{PhysRevA.82.012321}). This gives the following bound:
\bes
\begin{align}
|\Tr(\dot{\Pi}_0\Pi_0\rho\Pi_0^\perp)| &= |\Tr(S\dot{H}\Pi_0\rho\Pi_0^\perp)| \\
& \leq  \|S\dot{H}\| \|\Pi_0\rho\Pi_0^\perp\|_1 \\
& \leq  \frac{\|\dot{H}\|}{\Delta(H)}\|\Pi_0\rho\Pi_0^\perp\|_1\ ,
\end{align}
\ees
where $\|\cdot\|$ is the operator norm (largest singular value) and $\|\cdot\|_1$ is the trace-norm, and we used the inequality $|\Tr(AB)| \leq \|A\|\| B^\dagger\|_1$ (operator norm of $A$ times trace norm of $B^\dagger$)~\cite{Bhatia:book}.

How tightly can we bound the $\|\Pi_0^\perp\rho\Pi_0\|_1$ factor in general? As mentioned above, for the time-independent case and in the weak coupling limit this quantity decays rapidly due to decoherence between eigenstates. But in the time-dependent case the best we can do in general is the following.
Let $\rho^{\mathrm{ad}}$ be the solution of of the Davies-Lindblad master equation in the adiabatic limit. This is known to be the Gibbs state~\cite{Venuti:2015kq}, which is diagonal in the energy eigenbasis. Hence $\Pi_0^\perp\rho^{\mathrm{ad}}\Pi_0 = 0$, and using $\|\Pi_0\| = \|\Pi_0^\perp\|=1$:
\bes
\label{eq:B2}
\begin{align}
\|\Pi_0^\perp\rho\Pi_0\|_1 & = \|\Pi_0^\perp(\rho-\rho^{\mathrm{ad}})\Pi_0\|_1 \\
&\leq \|\Pi_0^\perp\| \|\rho-\rho^{\mathrm{ad}}\|_1\|\Pi_0\| \leq C/t_f\ ,
\label{eq:ad-th}
\end{align}
\ees
where we used the adiabatic theorem for open systems~\cite{Venuti:2015kq} for the last inequality. The $t_f$-independent constant $C$ is given by Eq.~(4) of ~\cite{Venuti:2015kq} and depends on a power of $ 1/\Delta(\mathcal{L})$ where $\mathcal{L}$ is the Lindbladian from Eq.~\eqref{eq:ME} and $\Delta(\mathcal{L})$ is the minimal gap from the zero eigenvalue of $\mathcal{L}(s)$ along the evolution from $s=0$ to $s=1$, where $s=t/t_f$. 

The bound~\eqref{eq:ad-th} can be tightened using boundary cancellation methods, and improved to $C_v/t_f^{v+1}$, where $v$ is the number of vanishing derivatives of $H(t)$ at  $t_f$, and the bounded, $t_f$-independent constants $C_v$ are given in Eq.~(11b) of Ref.~\cite{Venuti:2018aa}. It is important to note that $C_v$ does depend on the system size $n$, a point we return to below. Thus, altogether we have:
\beq
|\Tr[\dot{\Pi}_0\Pi_0\rho\Pi_0^\perp]| \leq \frac{\|\dot{H}\|C_v}{\Delta(H)t_f^{v+1}}\ ,
\label{eq:B3}
\eeq 
where $\|\dot{H}\|$ and $\Delta(H)$ are maximized and minimized over the interval $[0,t_f]$, respectively.

\subsection{Computation of the diagonal term}

Now we consider the diagonal term  in Eq.~\eqref{eq:R2}:
\beq
\Tr(\Pi_0\dot{\rho}) = -i\Tr(\Pi_0[H+H_{\mathrm{LS}},\r] + \Tr(\Pi_0 D[\rho]) \ .
\label{eq:diag}
\eeq
Let us compute each term in turn.

\subsubsection{Computation of $\Tr(\Pi_0[H+H_{\mathrm{LS}},\r])$}

Let $H'\equiv H+H_{\mathrm{LS}}$. 
Recall that $[H,H_{\mathrm{LS}}]=0$ and $[\Pi_0,H']=0$.
Now note that $\Pi_0[H',\r] = [H',\Pi_0 \r]+[\Pi_0,H']\r$, so that:
\beq
\Tr(\Pi_0[H',\r]) = \Tr([\Pi_0,H']\r) = 0\ .
\label{eq:no-unit}
\eeq 
Thus there is no contribution from the unitary part.

\subsubsection{Computation of $\Tr(\Pi_0 D[\rho])$}

Eq.~\eqref{eq:diss} gives:
\bes
\label{eq:C3}
\begin{align} 
\label{eq:C3a}
\Tr(\Pi_0 D[\rho])& =\sum_{\omega\a} \gamma_{\alpha} (\omega)\bigg(\Tr[\Pi_0 F_{\a}(\omega) \rho(t)F_{\alpha}^{\dagger}(\omega)]   \\
\label{eq:C3b}
	&\qquad -\frac{1}{2} 
	\Tr[\Pi_0 \{F_{\alpha}^{\dagger}(\omega)F_{\a}(\omega),\rho\}] \bigg)\ .
\end{align}
\ees
The term in line~\eqref{eq:C3a} is:
\begin{align}
&\sum_\omega \gamma_{\alpha}(\omega)\Tr[\Pi_0 F_{\a}(\omega) \rho F_{\alpha}^{\dagger}(\omega)\Pi_0 ]  \notag \\
&\quad = \sum_\omega \gamma_{\alpha}(\omega)\sum_{\epsilon_{l'}-\epsilon_l=\omega}\sum_{\epsilon_{l'''}-\epsilon_{l''}=\omega}\notag \\
&\qquad \Tr[\Pi_0({\Pi_l F_{\a}\Pi_{l'}})\rho (\Pi_{l'''}F^\dagger_{\alpha}\Pi_{l''})\Pi_0] \notag \\
&\quad = \sum_\omega \gamma_{\alpha}(\omega)\sum_{\epsilon_{l'}=\omega+\epsilon_0}\sum_{\epsilon_{l'''}=\omega+\epsilon_{0}}\notag \\
&\qquad \Tr[({\Pi_0 F_{\a}\Pi_{l'}})\rho (\Pi_{l'''}F^\dagger_{\alpha}\Pi_0)] \notag \\
&\quad = \sum_{l} \gamma_{\alpha}(\epsilon_{l}-\epsilon_{0})
\Tr[{\Pi_0 F_{\a}\Pi_{l}}\rho\Pi_{l}F^\dagger_{\alpha}\Pi_0] \ . 
\end{align}


For the term in line~\eqref{eq:C3b}, we first note that:
\begin{align}
&\Pi_0 F_{\alpha}^{\dagger}(\omega)F_{\a}(\omega)\notag\\
&\quad = \sum_{\epsilon_{l'}-\epsilon_l=\omega}\sum_{\epsilon_{l'''}-\epsilon_{l''}=\omega} \Pi_0({\Pi_{l'''}F^\dagger_{\alpha}\Pi_{l''}})
({\Pi_l F_{\a}\Pi_{l'}}) \notag\\
&\quad =\sum_{\epsilon_{l'}-\epsilon_l=\omega}\sum_{\epsilon_{0}-\epsilon_{l}=\omega} {\Pi_{0}F^\dagger_{\alpha}}
{\Pi_l F_{\a}\Pi_{l'}}\notag\\
&\quad =\sum_{\epsilon_{l}=\epsilon_0-\omega} {\Pi_{0}F^\dagger_{\alpha}}
{\Pi_l F_{\a}\Pi_{0}}\ .
\end{align}
An identical result holds for $ F_{\alpha}^{\dagger}(\omega)F_{\a}(\omega)\Pi_0$.
Using this, the term in line~\eqref{eq:C3b} becomes $-\sum_{\l}\gamma_{\alpha}(\e_0-\e_l)
	\Tr[{\Pi_{0} \rho \Pi_{0}F^\dagger_{\alpha}}{\Pi_l F_{\a}}]$.
Therefore, after using the KMS condition to write $\gamma_{\alpha}(\e_0-\e_l)=e^{-\b (\e_l -\e_0)}\gamma_{\alpha}(\e_l-\e_0)$:
\begin{align}
\label{eq:rate2}
\Tr(\Pi_0 D[\rho]) &= \sum_{\alpha}\sum_{l>0}\g_{\alpha}(\e_l -\e_0)\left( \Tr[{ \rho \Pi_{l}F^\dagger_{\alpha}\Pi_0 F_{\a}\Pi_{l}}]\right. \notag \\
& \left. \qquad -  e^{-\b (\e_l -\e_0)} \Tr[\rho \Pi_0 F_{\alpha}^{\dagger} \Pi_l F_{\a}  \Pi_0 ] \right) \ .
\end{align}
These terms represent opposite processes: the first represents relaxation into $\Pi_0$, the second represents excitation out of $\Pi_0$. When just the initial excitation rate is accounted for, only the excitation term appears~\cite{Marvian:2017aa}.

\subsubsection{The case without degeneracies: Pauli master equation}

In the absence of any degeneracies the projectors are all rank $1$, i.e., $\Pi_l = \ketbra{\e_l}$ and we can simplify Eq.~\eqref{eq:rate2} by factoring out the populations 
\beq
p_l = \Pi_l \rho \Pi_l\ .
\eeq
This yields:
\bes
\label{eq:rate2a}
\begin{align}
\Tr(\Pi_0 D[\rho]) &= \sum_{\alpha}\sum_{l>0}\g_{\alpha}(\e_l -\e_0)p_l \left( \Tr[{ \Pi_{l}F^\dagger_{\alpha}\Pi_0 F_{\a}}]\right. \notag \\
& \left. \qquad -  e^{-\b (\e_l -\e_0)} p_0 \Tr[\Pi_0 F_{\alpha}^{\dagger} \Pi_l F_{\a}] \right) \\
& = \sum_{l>0} W_{0l}p_l - W_{l0}p_0 \\
\end{align}
\ees
where we used the KMS condition and defined the Markov transition matrix (whose elements are positive):
\beq
W_{l'l} = \sum_{\alpha} \g_{\alpha}(\e_l -\e_{l'}) \Tr[{ \Pi_{l}F^\dagger_{\alpha}\Pi_{l'} F_{\a}}]\ .
\label{eq:MTM}
\eeq
In this case $\Tr(\Pi_0 \dot\rho) = \Tr(\Pi_0 D[\rho])$ is simply the Pauli master equation, expressing repopulation of the ground state $\ket{\e_0}$ with transition rate $W_{0l}$, and depopulation with transition rate $W_{l0} = e^{-\b (\e_l -\e_0)} W_{0l}$ (detailed balance). Both positivity and the detailed balance conditions are proved in Appendix~\ref{app:A}.

\section{Excitation rate reduction using an error detecting code}
\label{sec:3}

\subsection{The encoded Hamiltonian}

\begin{figure*}[t]
\includegraphics[width=0.8\textwidth]{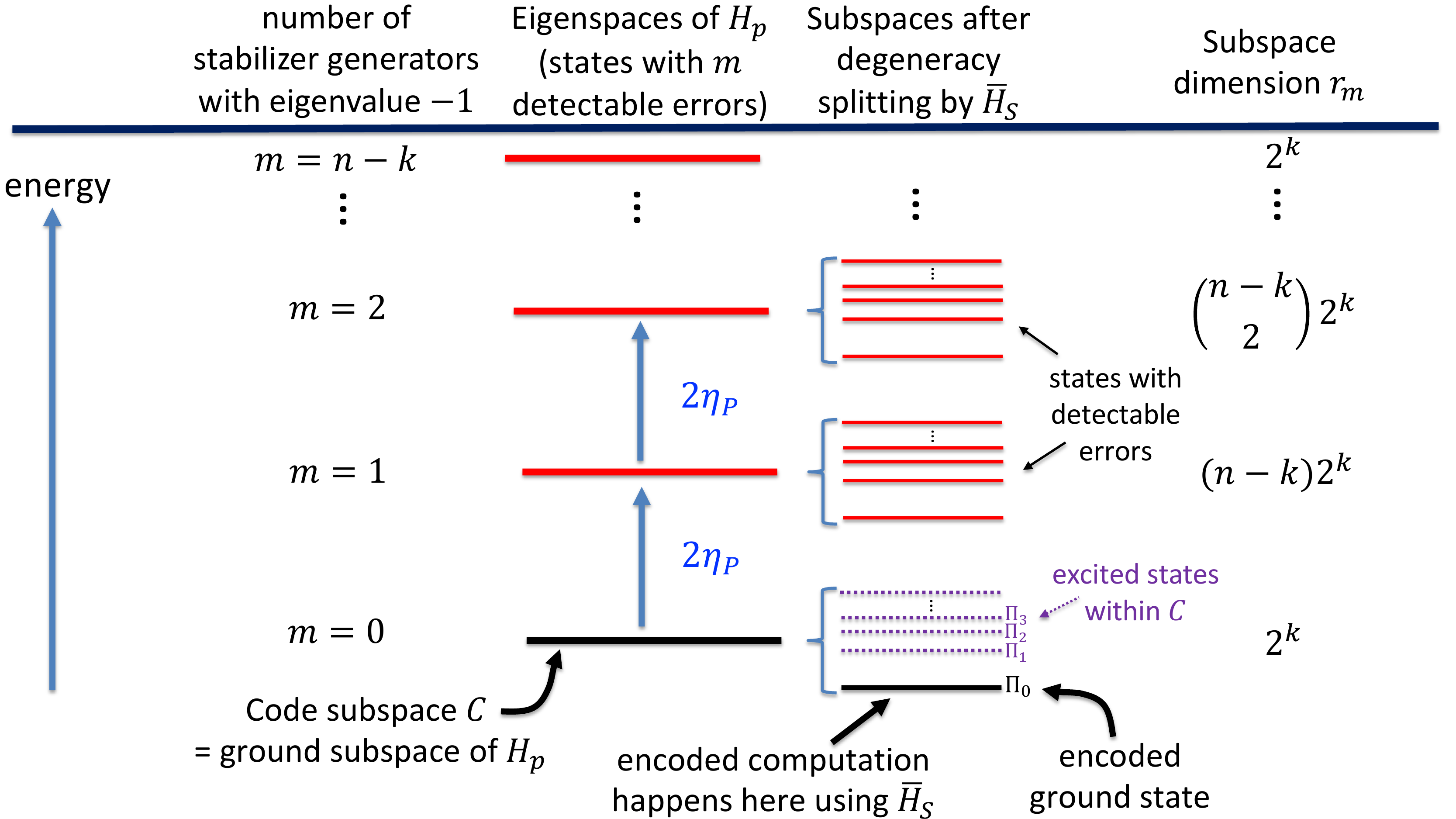}
\caption{(color online) Subspaces and energy level structure. Before the introduction of $\bar{H}_S$ the Hilbert space splits into the $n-k+1$ eigenspaces of the penalty Hamiltonian $H_\mrp$, indexed by the number of detected errors $m$. These subspaces are $\binom{n-k}{m}2^k$-dimensional and are separated by gaps of $2\eta_\mrp$. The lowest ($m=0$) is the code subspace $\mc{C}$. Transitions between different eigenspaces of $H_\mrp$ are suppressed by these gaps. After $\bar{H}_S$ is introduced the degeneracy of each such subspace is split. The encoded adiabatic computation takes place in the encoded ground state, the lowest energy state in $\mc{C}$. Diabatic evolution or undetected errors can cause transitions inside the code space, which are not protected against by the penalty Hamiltonian.}
\label{fig:subspace-code-diagram}
\end{figure*}

We now choose a code $C$ that can detect all the errors (system operators) $F_\a$~\cite{Knill:1997kx}:
\beq
\forall \a: \, P_C F_\alpha P_C=0\ ,
\label{eq:err-det}
\eeq
where $P_C$ projects onto the code space.
More explicitly, we choose $C$ to be an $[[n,k,d]]$ stabilizer code, where the number of physical and logical qubits is $n$ and $k$, respectively, and $d$ is the code distance ($d\geq 2$).
Let us denote the stabilizer generators by $\{S_i\}_{i=1}^{n-k}$. This partitions the Hilbert space into $2^{n-k}$ stabilizer syndrome subspaces, each of which is $2^k$-dimensional. Each such subspace is defined by a particular ordered assignment of $\pm 1$ eigenvalues of the stabilizer generators~\cite{Gottesman:1996fk}.

We construct a penalty Hamiltonian $H_{\mrp}$ by summing the stabilizer generators~\cite{jordan2006error}: 
\beq
H_{\mrp} = -\sum_{i=1}^{n-k}S_i\ . 
\eeq
The eigenvalues of $H_{\mrp}$ are 
\beq
\xi_m=-(n-k)+2m\ , \quad m=0,\dots,n-k\ ,
\label{eq:xi_m}
\eeq 
with corresponding $r_m = \binom{n-k}{m}2^k$-dimensional eigensubspaces having the property that exactly $m$ of the stabilizer generators have eigenvalue $-1$. Note that $\sum_{m=0}^{n-k} r_m = 2^k \sum_{m=0}^{n-k} \binom{n-k}{m} = 2^n$, as required. 

We define the codespace $C$ as usual as the linear span of the simultaneous eigenstates of all the stabilizer generators with eigenvalue $+1$~\cite{Gottesman:1996fk}. Therefore the codespace is the ground subspace of $H_{\mrp}$. 
The codespace ($m=0$) is $r_0=2^k$-dimensional, supporting $k$ logical qubits.
The code therefore has $k$ logical $X$ and $k$ logical $Z$ operators denoted $\bar{X}_i$ and $\bar{Z}_i$, respectively, and their products form logical $XX$ and $ZZ$ operators. We encode the system Hamiltonian $H_S(t)$ using these logical operators. 
The following encoded Hamiltonian is universal for AQC~\cite{Biamonte:07}:
\begin{align}
\bar{H}_S(t) &= \sum_{i=1}^k h_i^x(t)\bar{X}_i + h_i^z(t)\bar{Z}_i \notag \\
&\qquad + \sum_{i<j} J^x_{ij}(t) \bar{X}_i\bar{X}_j + J^z_{ij}(t)\bar{Z}_i\bar{Z}_j\ ,
\end{align}
so we may assume this form without loss of generality. The entire time-dependence is in the parameters $\{h_i^x(t),h_i^z(t),J^x_{ij}(t),J^z_{ij}(t)\}$.

The total system Hamiltonian is then
\ba
H(t)= \eta_{\mrp} H_{\mrp} + \bar{H}_S(t) \ ,
\label{eq:H(t)}
\ea 
where the dimensionless quantity $\eta_{\mrp}>0$ quantifies the strength of the energy penalty. Every error detected by the code anticommutes with at least one stabilizer generator [this is equivalent to Eq.~\eqref{eq:err-det}]~\cite{Gottesman:1996fk}, so every such error ``pays" an energy penalty equal to the number of anticommuting generators times $\eta_{\mrp}$.
Adding the encoded system Hamiltonian $\bar{H}_S(t)$ to $\eta_{\mrp} H_{\mrp}$ splits the $2^k$-fold degeneracy of the code space. The ground state of $H(t)$ is the encoded ground state which the error suppression scheme is designed to protect. The protection is against bath-induced errors that excite the system out of the code space ($m=0 \mapsto m>0$), but not against errors that induce transitions inside the code space ($m=0 \mapsto m=0$). The latter can be either logical errors due to coupling of the bath to system operators with weight $\geq d$, or due to diabatic transitions arising from non-adiabatic evolution. These considerations, as well as additional ones discussed below, are illustrated in Fig.~\ref{fig:subspace-code-diagram}.

Since by construction $[H_{\mrp},\bar{H}_S(t)]=0$ $\forall t$, it follows that we may express the two Hamiltonians in terms of the same set of time-dependent eigenprojectors:
\beq
H_{\mrp} = \sum_{l\ge 0} \xi_l \Pi_l(t)\ , \qquad  \bar{H}_S(t) = \sum_{l\ge 0} \bar{\o}_l(t) \Pi_l(t)\ ,
\label{eq:proj}
\eeq
where the $\{\xi_l\}_l$ and $\{\bar{\o}_l(t)\}_l$ are, respectively, the eigenvalues of $H_{\mrp}$ and $\bar{H}_S(t)$. Since $\bar{H}_S(t)$ breaks the degeneracy of $H_{\mrp}$, the index $l$ ranges over a set of values that is at least as large as that of the index $m$ in Eq.~\eqref{eq:xi_m}. I.e., unlike in Eq.~\eqref{eq:xi_m} where $m\neq m' \Longrightarrow \xi_m\neq \xi_{m'}$, the index $l$ may repeat certain eigenvalues of $H_{\mrp}$, and $\xi_l= \xi_{l'}$ is possible when $l\neq l'$.%
\footnote{
It may seem surprising that the time-independent $H_{\mrp}$ can be expressed in terms of a linear combination of time-dependent projectors. To see this explicitly, let $\{\xi_m,\{\ket{v_j^{(m)}}\}_{j=1}^{r_m}\}_m$ denote the time-independent eigenvalues and corresponding eigenvectors of  $H_{\mrp}$, where $r_m$ is the degeneracy of $\xi_m$ [recall Eq.~\eqref{eq:xi_m}]. Since $[H_{\mrp},\bar{H}_S(t)]=0$, $\bar{H}_S(t)$ preserves the eigenspaces of $H_{\mrp}$ but may break the degeneracy within each such subspace, and the eigenvectors of $\bar{H}_S(t)$ are time-dependent linear combinations of the eigenvectors of $H_{\mrp}$. I.e., we can write the eigenvectors of $\bar{H}_S(t)$ as $\{\{\ket{w_k^{(l)}(t)} = \sum_{j} a_{jk}^{(l)}(t) \ket{v_j^{(l)}}\}_{k}\}_l$, with corresponding time-dependent eigenvalues $\{\bar{\o}_l(t)\}_l$. We can trivially check that the time-dependent vectors $\{\{\ket{w_k^{(l)}(t)}\}_{k}\}_l$ are also eigenvectors of $H_{\mrp}$ with eigenvalues $\{\xi_l\}_l$. Namely, $H_{\mrp}\ket{w_k^{(l)}(t)} = \sum_j a_{jk}^{(l)}(t) H_{\mrp}\ket{v_j^{(l)}} =  \sum_j a_{jk}^{(l)}(t) \xi_l\ket{v_j^{(l)}} = \xi_l \ket{w_k^{(l)}(t)}$. Therefore we may express $H_{\mrp}$ as in Eq.~\eqref{eq:proj}, with the projectors explicitly identified as $\Pi_l(t) = \sum_{k} \ketbra{w_k^{(l)}(t)}$.
}

We may now choose the $\Pi_l$'s of Eq.~\eqref{eq:H} as the same eigenprojectors and write the eigendecomposition of $H(t)$ as
\beq
H(t) = \sum_{l\geq 0} \e_l(t) \Pi_l(t)\ ,\quad \epsilon_l(t) = \eta_{\mrp} \xi_l +\bar{\o}_l(t) \ .
\label{eq:e_l}
\eeq

As mentioned above, the codespace is the ground space of $H_{\mrp}$. We may associate a corresponding projection operator 
\beq
P_C=\sum_{l \in C}{\Pi_l}\ .
\label{eq:P_C}
\eeq 
The fact that we have a sum over $l$ is due to the breaking of the degeneracy of the codespace by $\bar{H}_S(t)$; the sum is over all $l$ required to span the $2^k$-dimensional codespace. Eq.~\eqref{eq:e_l} tells us the ground subspace projector of $H(t)$ is $\Pi_0$, 
i.e., the ground space of $H(t)$ is a subspace of the code space.

\subsection{Modified excitation rate after encoding}

We assume that the initial state, which is now encoded into $C$  and evolves according to $H(t)$ of Eq.~\eqref{eq:H(t)}, belongs to $\Pi_0$, i.e., again $\rho(0)=\Pi_0\rho(0)\Pi_0$.
This means that $p_\perp(0)=0$ in Eq.~\eqref{eq:p-perp}, and we can focus exclusively on the excitation rate $R(t)$. Let us collect the results from above [Eqs.~\eqref{eq:R2}, \eqref{eq:B3}-\eqref{eq:no-unit},~\eqref{eq:rate2}], as follows. 

First, we note that the off-diagonal term $Q(t) \equiv \Tr(\dot{\Pi}_0 \Pi_0\rho\Pi_0^\perp) + \textrm{c.c.}$ satisfies the bound given in Eq.~\eqref{eq:B3}:
\beq
|Q(t)| \leq \frac{2\|\dot{H}\|C_v}{\Delta(H)t_f^{v+1}}\ , \quad t\in[0,t_f] \ .
\label{eq:|Q|}
\eeq 
Next:
\begin{subequations}
\begin{align}
R(t) - Q(t) &= \Tr({\Pi_0}\dot{\rho}) \\
&=  -i\Tr(\Pi_0[H+H_{\mathrm{LS}},\r] + \Tr(\Pi_0 D[\rho]) \\
&= \sum_{\alpha}\sum_{l>0}\g_{\alpha}(\e_l -\e_0)\left( \Tr[{ \rho(t) \Pi_{l}F^\dagger_{\alpha}\Pi_0 F_{\a}\Pi_{l}}]\right. \notag \\
& \left. \qquad -  e^{-\b (\e_l -\e_0)} \Tr[\rho(t) \Pi_0 F_{\alpha}^{\dagger} \Pi_l F_{\a}  \Pi_0 ] \right) \ .
\end{align}
\end{subequations}
Thanks to the error detection properties of the code [Eq.~\eqref{eq:err-det}] we have 
\begin{align}
\forall l \in C: \Pi_0 F_\alpha \Pi_l= 0\ ,
\end{align}
so that the sum over $l$ reduces to a sum only over terms not in the codespace:
%
%
%
%
%
\bes
\begin{align}
\label{eq:purity-loss2}
&R(t)  = \sum_{\a}\sum_{l\in C^{\perp}}\g_{\alpha}(\e_l -\e_0) (m_{\a}^{l,0}
-e^{-\b (\e_l -\e_0)} m_{\a}^{0,l} ) + Q(t)\\
& m_{\a}^{a,b} \equiv \Tr[{ \rho(t) \Pi_{a}F^\dagger_{\alpha}\Pi_b F_{\a}\Pi_{a}}]\ .
\end{align}
\ees
The term $m_{\a}^{0,l}$ represents the excitation out of the codespace associated with error operator $F_\a$, while $m_{\a}^{l,0}$, represents the corresponding relaxation into the codespace.

\subsection{The non-codespace population is exponentially suppressed by the energy penalty}
Next, let us show that for reasonable models of the bath the excitation rate is exponentially suppressed with increasing energy penalty $\eta_{\mrp}$. 

First, repeating the argument given in Ref.~\cite{Marvian:2017aa}, let $\Pi_l$ denote an eigenprojector of $H$ with energy $\epsilon_l = \Tr[\Pi_l H]$. Recall that the $\Pi_l$'s are simultaneous eigenprojectors of $\bar{H}_S$ and $H_{\mrp}$ as well. We have $\forall l \in C^{\perp}$:  
\begin{align}
\e_l -\e_0 &=\Tr[\Pi_l(\bar{H}_S +\eta_{\mrp} H_{\mrp})]-\Tr[\Pi_0(\bar{H}_S +\eta_{\mrp} H_{\mrp})]\notag \\
&=\Tr[(\Pi_l-\Pi_0)\bar{H}_S] +\eta_{\mrp} \Tr[(\Pi_l -\Pi_0)H_{\mrp}] \notag \\
&\geq \eta_{\mrp}g \ ,
\end{align}
where $g$ is the ground state gap of $H_{\mrp}$:
\beq
g \equiv \min_{l \in C^\perp}  \Tr[(\Pi_l -\Pi_0)H_{\mrp}]\ .
\eeq 
When $H_{\mrp}$ is a sum of commuting terms, as is true for the stabilizer construction we consider here, the gap $g$ is guaranteed to be a constant~\cite{Bravyi:2003tx}.

Next, note that, using the spectral decomposition $\rho(t) = \sum_i \lambda_i \ketbra{i}$, both the excitation and relaxation terms in Eq.~\eqref{eq:purity-loss2} are positive: 
\beq
m_{\a}^{a,b} = \sum_i \lambda_i \| \Pi_b F_{\a} \Pi_a\ket{i}\|^2\geq 0\ .
\label{eq:m-pos}
\eeq
Using Eqs.~\eqref{eq:p-perp} and~\eqref{eq:purity-loss2}, the non-codespace population is \begin{align}
\label{eq:pop-perp}
& p_\perp(t) = -\int_0^t R(t')dt'  =\int_0^t Q(t')dt'+  \notag \\
& \quad \sum_{\alpha}\sum_{l\in C^{\perp}}\int_0^t \g_{\alpha}(\e_l -\e_0) \bigg( e^{-\b (\e_l -\e_0)} m_{\a}^{0,l}-  m_{\a}^{l,0}\bigg)dt' \notag \ . 
\end{align}
We may replace $e^{-\b (\e_l -\e_0)}$ by the upper bound $e^{-\b \eta_{\mrp}g}$, and further increase the RHS by removing the relaxation term, since it is positive.
Hence:
\bes
\label{eq:47}
\begin{align}
p_\perp(t) &\leq \sum_{\alpha}\sum_{l\in C^{\perp}}\int_0^t  \g_{\alpha}[\e_l(t') -\e_0(t')] e^{-\b \eta_{\mrp}g} m_{\a}^{0,l}(t')dt' \notag \\
&\qquad + \int_0^t |Q(t')|dt' \\
&\leq t \tilde{\g}_{\max} e^{-\b \eta_{\mrp}g} \sum_{\alpha}\sum_{l\in C^{\perp}} \tilde{m}_{\a}^{0,l} + t |Q|\ ,
\label{eq:pop-perp2}
\end{align}
\ees
where we defined
\bes
\begin{align}
\label{eq:gmax}
{\tilde{\g}_{\max}} &{\equiv} \max_{l \in C^{\perp}, \a, t\in[0,t_f]}\gamma_{\alpha}[\e_l(t)-\e_0(t)]\\ 
\label{eq:gamma_max}
&= \max_{l \in C^{\perp}, \a}\gamma_{\alpha}(\bar{\o}_l + \eta_{\mrp} \xi_l-\e_0)\\
\tilde{m}_{\a}^{0,l} &\equiv \max_{t\in[0,t_f]} {m}_{\a}^{0,l}(t)\\
|Q| &\equiv \max_{t\in[0,t_f]} |Q(t)| \ ,
\end{align}
\ees
and used the fact that we already know that both ${\tilde{\g}_{\max}}$ and $\tilde{m}_{\a}^{0,l} $ are positive. To obtain line~\eqref{eq:gamma_max} we used Eq.~\eqref{eq:e_l}. Note the appearance of the factor $t$ due to the integration in Eq.~\eqref{eq:47}. This factor will play an important role in our final upper bound considerations below, and was absent from the initial-time-only considerations of Refs.~\cite{jordan2006error,Marvian:2017aa}. 

The bound on $|p_\perp|$ depends on ${\tilde{\g}_{\max}}$.
To ensure a non-trivial bound this quantity has to be finite, which is a natural assumption. 
%
We also assume that $\g(\o)$ is a polynomial (or any subexponential) function of $\o$ for $\o>0$; this too is an assumption that is compatible with all commonly used bath models~\cite{Breuer:2002}. 
Therefore $\tilde{\g}_{\max}\sim \mathrm{poly}(\eta_{\mrp})$.

What remains to be shown is that the sum over all non-code states---which might appear to involve exponentially many terms---does not spoil this conclusion. Now, since we already showed [Eq.~\eqref{eq:m-pos}] that each $m_{\a}^{0,l} \geq 0$, we have:
\beq
\sum_{\alpha}\sum_{l\in C^{\perp}} \tilde{m}_{\a}^{0,l} \leq \sum_{\alpha,l} \tilde{m}_{\a}^{0,l} =  \max_{t\in[0,t_f]} \Tr[\Pi_{0}\rho \Pi_{0}\sum_{\alpha} F^\dagger_{\alpha}F_{\a}]\ ,
\label{eq:poly-growth}
\eeq
where we used $\sum_l \Pi_l =I$.
Using the inequality $|\Tr(AB)| \leq \|A\|\| B^\dagger\|_1$ again, we have
\bes
\begin{align}
\Tr[\Pi_0\rho\Pi_0 F_{\a}^{\dagger} F_{\a} ] &= |\Tr[\Pi_0\rho\Pi_0 F_{\a}^{\dagger} F_{\a}] | \\
&\leq \|F_{\a}^{\dagger} F_{\a}\| \leq \|F_{\a}\|^2
\ ,
\end{align}
\ees
since $\|\Pi_0\rho\Pi_0\|_1\leq 1$.
Thus:
\beq
\sum_{\alpha}\sum_{l\in C^{\perp}} \tilde{m}_{\a}^{0,l} \leq \max_{t\in[0,t_f]}\sum_{\alpha} \|F_{\a}\|^2\ .
\eeq
The sum over $\a$ contains a polynomial number of terms in $n$ due to our earlier assumption of $\ell$-local system operators in the system-bath interaction Hamiltonian $H_{SB}=\sum_{\alpha}{A_{\alpha} \otimes B_{\alpha}}$. Each $\|F_{\a}\|$ is itself a sum over $\mathrm{poly}(n)$ $n$-independent terms due to Eq.~\eqref{eq:F}.

Combining all this with Eqs.~\eqref{eq:|Q|} and \eqref{eq:pop-perp2} we thus conclude that
\bes
\label{eq:pop-perp-final}
\begin{align}
\label{eq:pop-perp-final1}
p_\perp(t_f) &\leq t_f \exp(-\b g \eta_{\mrp})\mathrm{poly}(\eta_{\mrp}) \mathrm{poly}(n)  \\
&\qquad + \frac{2\|\dot{H}\|C_v}{\Delta(H)t_f^{v}} \ .
\label{eq:pop-perp-final2}
\end{align}
\ees
Line~\eqref{eq:pop-perp-final1} states that the non-codespace population is exponentially suppressed in terms of the energy penalty, and grows at most linearly in time. This contribution to the bound is due to the error suppression strategy, and is similar to the result in Ref.~\cite{Marvian:2017aa}. The new aspect is the factor of $t_f$.
The dependence of $t_f$ on problem size $n$ is dictated by the adiabatic theorem: $t_f$ must scale as a small inverse power ($2$ or $3$) of the minimum gap $\Delta(H)$ encountered along the evolution~\cite{Jansen:07,lidar:102106}. The same condition also ensures that the denominator in line~\eqref{eq:pop-perp-final2} grows with $n$. This line arises purely due to diabatic transitions, which cannot be suppressed using the energy penalty or encoding. Note that one factor of $t_f$ present in Eq.~\eqref{eq:|Q|} cancelled after integration, so that we must enforce $v\geq 1$.

To ensure that the entire contribution of line~\eqref{eq:pop-perp-final2} decreases with $n$, the number of vanishing derivatives $v$ at $t=t_f$ must be sufficiently large to overcome both the scaling of $\|\dot{H}\|$ with $n$ (which is at most quadratic, in the case of all-to-all interactions), and the scaling of $C_v$ with $n$. The latter depends on powers of both $\Delta(H)$ and $\Delta(\mathcal{L})$, since it is known that in the Davies-Lindblad adiabatic master equation case and in the presence of a thermal bath satisfying the KMS condition, the constant $C$ appearing in the adiabatic theorem bound without assuming any boundary conditions [Eq.~\eqref{eq:ad-th}] satisfies $C = O[\|\mathcal{L}'\|/\Delta^2(\mathcal{L})]$ (where prime denotes differentiation with respect to $s$)~\cite{Venuti:2017aa}.

Therefore, \emph{assuming that all inverse gap dependencies are polynomial in $n$}, Eq.~\eqref{eq:pop-perp-final} implies that as long as $t_f\sim \mathrm{poly}(n)$ (with an appropriately high degree) then \emph{one only needs to increase the strength of the energy penalty, $\eta_{\mrp}$, logarithmically in $n$}, at any fixed inverse temperature $\b$. 

To see explicitly why, let $ \mathrm{poly}(n) \sim n^p$, $t_f \sim n^q$, and $\eta_{\mrp} \sim r\log(n)$. Then
\begin{subequations}
\begin{align}
t_f \exp(-\b g \eta_{\mrp}) \mathrm{poly}(n) & \sim \exp[\log(n^{-\b g r})] n^{p+q} \\
&= n^{p+q-\b g r} \ ,
\end{align}
\end{subequations}
so that if $r > (p+q)/(\b g)$ then this guarantees that line~\eqref{eq:pop-perp-final1} decreases (polynomially) in the system size $n$. The $\mathrm{poly}(\eta_{\mrp})$ factor  does not change this conclusion since it is a polynomial in $\log(n)$.

If, on the other hand, the inverse gap dependencies are exponential in $n$ then we need $t_f\sim \exp(n)$, and then $\eta_{\mrp}$ must grow at least linearly in $n$ in order for error suppression to be effective, which is unacceptable: the same suppression could be achieved simply by scaling up all the coupling constants linearly without incurring the cost of encoding.

The main conclusion reported in Refs.~\cite{jordan2006error,Marvian:2017aa} therefore remains valid for arbitrary evolution times, namely, that by using error detecting codes built on commuting Hamiltonians (for which $g$ is constant), for physically plausible Markovian models, a logarithmically increasing energy penalty strength suffices for error suppression. The main new caveat is that the proof holds for problems with polynomially small gaps, but not for problems with gaps that decrease superpolynomially.

\section{Summary and Discussion}
\label{sec:conc}

In this paper we studied error suppression for Markovian models. These are interesting, despite the fact that general results of a similar nature have already been established for non-Markovian models~\cite{Bookatz:2014uq,Marvian:2016aa}, since they are widely used~\cite{Breuer:2002}, and moreover decay in these models is always exponential~\cite{Alicki:87}. 
In this sense error suppression in the non-Markovian case is less challenging than for Markovian AQC, since in the latter case one cannot rely on the use of non-Markovian recurrences, as is commonly done in error suppression techniques such as dynamical decoupling~\cite{PhysRevLett.100.160506,PhysRevA.86.042333,Ganti:13} or the Zeno effect~\cite{PhysRevLett.108.080501}.

Ref.~\cite{Marvian:2017aa} generalized the JFS result~\cite{jordan2006error}, that it suffices to increase the energy penalty logarithmically with system size in order to protect AQC against excitations out of the ground state, to general Markovian dynamics and mixed states. However, these earlier results were only valid for the \emph{initial} excitation rate, and the natural question of whether they generalize to arbitrary evolution times was left open. Here we settled this question in the affirmative, under the assumption that the problem gap is at most polynomially small in the problem size. While it seems unlikely, we did not rule out the possibility that this method of error suppression could be adapted to work even for problems with an exponentially small gap. Still, the present result establishes that one of the main failure modes of AQC can be overcome under plausible physical assumptions for problems for which AQC is efficient. 

We emphasize that our results require the ability to encode both the final and the initial Hamiltonian. Therefore they do not apply to transverse field implementations of quantum annealing, where only the final Hamiltonian can be encoded~\cite{PAL:13,vinci2015nested}. A further caveat is that it is known that for penalty Hamiltonians comprising a sum of commuting Hamiltonians, as is the case here, at least $3$-local interactions are required~\cite{Marvian:2014nr}, and for stabilizer subspace codes at least $4$-local interaction are required for universality~\cite{jordan2006error}. However, we expect that our entire construction will generalize straightforwardly to the stabilizer subsystem setting~\cite{poulin_stabilizer_2005}, where the penalty Hamiltonian becomes a sum over the $2$-local generators of the (non-Abelian) gauge group of the code~\cite{Jiang:2015kx,Marvian-Lidar:16}. We expect this to improve upon the locality of the construction presented here as well. The generalization of our results to the subsystem code case is an important problem left for future work.

\acknowledgments
This work benefited greatly from numerous constructive discussions with Milad Marvian, who carefully read and improved the manuscript. Thanks also to Paolo Zanardi and Lorenzo Campos Venuti for insightful comments. The research is based upon work (partially) supported by the Office of
the Director of National Intelligence (ODNI), Intelligence Advanced
Research Projects Activity (IARPA), via the U.S. Army Research Office
contract W911NF-17-C-0050. The views and conclusions contained herein are
those of the authors and should not be interpreted as necessarily
representing the official policies or endorsements, either expressed or
implied, of the ODNI, IARPA, or the U.S. Government. The U.S. Government
is authorized to reproduce and distribute reprints for Governmental
purposes notwithstanding any copyright annotation thereon.

\appendix

\section{Properties of the Markov transition matrix}
\label{app:A}

We defined the Markov transition matrix in Eq.~\eqref{eq:MTM}: $W_{l'l} = \sum_{\alpha} \g_{\alpha}(\e_l -\e_{l'}) \Tr[{ \Pi_{l}F^\dagger_{\alpha}\Pi_{l'} F_{\a}}]$.

Positivity of the matrix elements can be seen by recalling Eq.~\eqref{eq:F} and that the $A_{\a}$ are Hermitian. Then:
\beq
\bra{\e_l}F^\dagger_{\alpha} \ket{\e_{l'}}^* = \sum_{\a'} u^*_{\a'\a} \bra{\e_{l'}}A_{\a'} \ket{\e_{l}} = \bra{\e_{l'}}F_{\alpha} \ket{\e_{l}}\ ,
\eeq 
so that:
\begin{subequations}
\begin{align}
W_{l'l} &= \sum_{\alpha} \g_{\alpha}(\e_l -\e_{l'}) \bra{\e_l}F^\dagger_{\alpha}\ketbra{\e_{l'}} F_{\a}\ket{\e_l} \\
&= \sum_{\alpha} \g_{\alpha}(\e_l -\e_{l'}) |\bra{\e_{l'}} F_{\a}\ket{\e_l}|^2 \geq 0\ .
\end{align}
\end{subequations}

To prove that detailed balance holds, let us write:
%
\bes
\begin{align}
W_{l'l} &= \sum_{\alpha} \g_{\alpha}(\e_l -\e_{l'}) \bra{\e_l}F^\dagger_{\alpha}\ketbra{\e_{l'}} F_{\a}\ket{\e_l} \\
& = \sum_{\alpha} \g_{\alpha}(\e_l -\e_{l'}) \sum_{\a'}u^*_{\a'\a} A_{l'l,\a'}\sum_{\a''} u_{\a''\a}A_{ll'\a''} \\
&= \sum_{\a'\a''} \g_{\a''\a'}(\e_l -\e_{l'}) A_{l'l,\a'} A_{ll',\a''} 
\label{eq:Wl'l}
\end{align}
\ees
where $ \g_{\a''\a'}(\e_l -\e_{l'}) = \sum_{\a} u_{\a''\a} \g_{\a}(\e_l -\e_{l'}) (u^\dag)_{\a\a'}$ and $A_{l'l,\a} \equiv \bra{\e_{l'}} A_{\a} \ket{\e_l}$. Thus
\begin{align}
W_{ll'} &= \sum_{\a'\a''} \g_{\a''\a'}(\e_{l'} -\e_{l}) A_{ll',\a'} A_{l'l,\a''} 
\end{align}
But using the the KMS condition with $\e_{l'}>\e_l$, we have from Eq.~\eqref{eq:Wl'l}
\begin{align}
W_{l'l} & = e^{-\b(\e_{l'} -\e_{l})}\sum_{\a'\a''}\g_{\a'\a''}(\e_{l'} -\e_{l})A_{l'l,\a'}A_{ll',\a''}\\
&=e^{-\b(\e_{l'} -\e_{l})}W_{ll'}\ .
\end{align}


%

\end{document}